\documentclass[aps,prd,twocolumn,showpacs,eqsecnum,nofootinbib]{revtex4}
\usepackage[dvips]{epsfig}
\usepackage{amssymb}
\usepackage{amsmath}
\usepackage{fancybox}

\begin{document}

\title{Do solar neutrinos decay?}
\author{John F. Beacom}
\author{Nicole F. Bell}
\affiliation{NASA/Fermilab Astrophysics Center, 
Fermi National Accelerator Laboratory, Batavia, IL 60510-0500, USA\\
{\tt beacom@fnal.gov, nfb@fnal.gov}}
\date{April 9, 2002}

\begin{abstract}
Despite the fact that the solar neutrino flux is now well-understood
in the context of matter-affected neutrino mixing, we find that it is
not yet possible to set a strong and model-independent bound on solar
neutrino decays.  If neutrinos decay into truly invisible particles,
the Earth-Sun baseline defines a lifetime limit of $\tau/m \agt
10^{-4}$ s/eV.  However, there are many possibilities which must be
excluded before such a bound can be established. There is an obvious
degeneracy between the neutrino lifetime and the mixing parameters.
More generally, one must also allow the possibility of active daughter
neutrinos and/or antineutrinos, which may partially conceal the
characteristic features of decay. Many of the most exotic
possibilities that presently complicate the extraction of a decay
bound will be removed if the KamLAND reactor antineutrino experiment
confirms the large-mixing angle solution to the solar neutrino problem
and measures the mixing parameters precisely.  Better experimental and
theoretical constraints on the $^8$B neutrino flux will also play a
key role, as will tighter bounds on absolute neutrino masses.  Though
the lifetime limit set by the solar flux is weak, it is still the
strongest direct limit on non-radiative neutrino decay.  Even so,
there is no guarantee (by about eight orders of magnitude) that
neutrinos from astrophysical sources such as a Galactic supernova or
distant Active Galactic Nuclei will not decay.
\end{abstract}

\pacs{13.35.Hb, 14.60.Pq, 26.65.+t    \hspace{2cm} FERMILAB-Pub-02/061-A}
\maketitle


\section{Introduction}

Since solar neutrinos have been detected with roughly the expected
flux, it appears that they do not decay over the 500 s $\times\ c$
distance to Earth.  Furthermore, neutrinos from SN1987a were also
detected in reasonable numbers on a much longer baseline of ($5 \times
10^{12}$ s) $\times\ c$.  Decay will deplete the flux of neutrinos of
energy $E$ and mass $m$ over a distance $L$ by the factor
\begin{equation}
\exp\left(-\frac{t}{\tau_{lab}}\right) = 
\exp\left(-\frac{L}{E} \times \frac{m}{\tau}\right)\,,
\end{equation}
where $\tau$ is the rest-frame lifetime and we use $c = 1$ units from
now on.  In Table~I, we list representative $\tau/m$ scales for
various neutrino sources.

\begin{table}[t]
\label{sources}
\caption{Representative scales for neutrino lifetimes, taken simply as
$\tau/m \sim L/E$.  The top three entries correspond to present data,
and the lower two to possible future data (after neutrinos of all
flavors have been observed from a Galactic supernova and neutrinos
from astrophysical sources like active galactic nuclei (AGN) have been
observed in km$^3$ detectors).}
\begin{tabular}{lcr}
Neutrino source & $L/E$               & $\tau/m$ (s/eV) \\
\hline\hline 
Accelerator     & 30 m/10 MeV         & $10^{-14}$ \\ 
Atmosphere      & $10^4$ km/300 MeV   & $10^{-10}$ \\ 
Sun             & 500 s/5 MeV        & $10^{-4}$ \\
\hline 
Supernova       & 10 kpc/10 MeV       & $10^5$ \\
AGN             & 100 Mpc/1 TeV       & $10^4$ \\ 
\hline\hline
\end{tabular}
\end{table}

In this paper, we critically assess what the best limits on neutrino
decay are.  We find that it is not yet possible to set
model-independent bounds, even for the well-measured solar neutrinos.
We discuss how decay limits can be improved in the future.

Though in the past neutrino decay was frequently discussed in terms of
{\it flavor} eigenstates, the lifetimes of neutrinos are only
well-defined for {\it mass} eigenstates (a flavor state does not have
a definite mass, lifetime, or magnetic moment).  Since we now know
that mixing angles are large, this distinction is essential.

Therefore, in considering the decay of neutrinos from the Sun and
SN1987a, one has to properly assess the mass eigenstate content of the
fluxes.  The SN1987a data can be reasonably explained by saying that
the expected flux of $\bar{\nu}_1$ made it to Earth and was detected
as $\bar{\nu}_e$ via the charged-current reaction $\bar{\nu}_e p
\rightarrow e^+ n$.
The $\bar{\nu}_\mu$ component would only have been detectable in
neutral-current reactions, and the SN1987a data are consistent with no
neutral-current events.  If $\nu_1$ is (as suggestively labeled) the
lightest mass eigenstate, then it would not be kinematically allowed
to decay, so that the decay limit from SN1987a would be meaningless.

Presently, the best explanation of the solar neutrino problem is large
mixing angle (LMA) Mikheyev-Smirnov-Wolfenstein (MSW) transformation
of $\nu_e$ to $\nu_\mu,\nu_\tau$.  Besides being the best
oscillation-parameter fit, LMA also provides a ``good'' fit to all of
the solar neutrino data in terms of an acceptable chi-squared.
Without the effects of oscillations, the solar neutrino flux is only
understood at the factor of two level.  Including the effects of
oscillations, the total flux of all flavors is better understood, and
the hope is that much more stringent decay limits can be derived.  A
similar approach was used to set the strongest direct neutrino
magnetic moment limit~\cite{magnetic}.

If LMA is the correct scenario, then solar neutrinos in the $5-15$ MeV
energy range are created as nearly pure $\nu_2$ mass eigenstates.
Furthermore, their propagation in the Sun is completely adiabatic, so
that they emerge as pure $\nu_2$ eigenstates, where $m_2 > m_1$.
Since the characteristic signature of decay is its energy dependence,
we restrict our attention to this energy range, over which
Super-Kamiokande (SK)~\cite{SK} and the Sudbury Neutrino Observatory
(SNO)~\cite{SNO} have measured energy spectra.  This data may thus be
used to search for the signatures of the kinematically allowed
$\nu_2 \rightarrow \nu_1 + X$
decay (the decay modes are discussed in detail below).  

While neutrino decay was an early proposed explanation of the solar
neutrino problem~\cite{solardecay,solardecay1}, in this paper we will
take the point of view that the LMA solution specifies the correct
basic picture and will consider $\nu_2 \rightarrow \nu_1 + X$ decay as
a perturbation, with the goal of placing a limit on $\tau/m$ (since
$m$ is unknown, only the quantity $\tau/m$ can be constrained).
Ongoing experiments will soon confirm or refute the LMA solution.

While there are strong limits on {\it radiative} neutrino
decays~\cite{RPP,raffelt}, in this paper we consider only the
so-called ``invisible'' decays, i.e., decays into possibly detectable
neutrinos (or antineutrinos) plus truly invisible particles, e.g.,
light scalar or pseudoscalar bosons.  For these modes, the lifetime
limits are very weak indeed; see Table~I.  There are also limits
from cosmology, the strongest of which use the cosmic microwave
background data~\cite{cosmolimits}.  The effect searched for arises
because of the transfer of nonrelativistic energy density in the
parent neutrinos ($m \agt 10$ eV) to relativistic energy density in
the daughter neutrinos ($m \simeq 0$).  Neither the large mass of the
parent neutrino nor the huge mass splitting is supported by present
data, so these cosmological limits on non-radiative neutrino decays
are inapplicable.

\begin{figure}[t]
\begin{center}
\epsfig{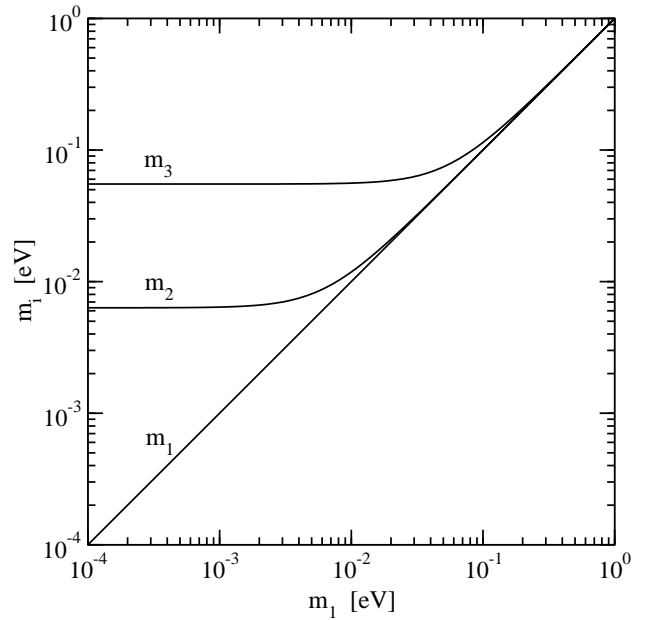}
\caption{\label{levels} The values of the neutrino mass eigenvalues 
as a function of
the unknown smallest mass $m_1$.  When $m_1$ is specified, the
measured solar and atmospheric $\delta m^2$ values fix $m_2$ and
$m_3$.  For an inverted hierarchy ($\delta m^2_{atm} < 0$), the two
upper eigenvalues are very nearly degenerate at the position of the
curve labeled $m_3$.  The present bound on neutrino mass from tritium
experiments is about 2 eV~\cite{tritium}; by the construction above,
it applies to all three mass eigenvalues.}
\end{center}
\end{figure}

While detection of daughter neutrinos or antineutrinos has been
considered in the literature, it has nearly always been with the
assumption that $m_1 \ll m_2$, and that the daughter neutrino carries
half the energy of its parent in a two-body decay.  These assumptions
are not generally valid.  Since oscillation experiments do not
determine the overall mass scale, the lightest mass eigenvalue $m_1$
is unknown.  However, for fixed $m_1$, the masses $m_2$ and $m_3$ are
determined by the measured mass-squared differences.  Thus
\begin{equation}
m_2 = \sqrt{m_2^2 - m_1^2 + m_1^2} = \sqrt{m_1^2 + \delta m^2_{sol}}\,,
\end{equation}
where solar-neutrino data give $\delta m^2_{sol} \simeq 4 \times
10^{-5}$ eV$^2$, and similarly,
\begin{equation}
m_3 = \sqrt{m_1^2 + \delta m^2_{sol} + \delta m^2_{atm}}\,,
\end{equation}
where atmospheric-neutrino data give $\delta m^2_{atm} \simeq 3 \times
10^{-3}$ eV$^2$.  All three masses are shown in Fig.~\ref{levels},
illustrating that the masses are nearly degenerate unless the overall
mass scale is tiny.  The accepted wisdom is that neutrino masses
should be strongly hierarchical, e.g., in see-saw models, proportional
to the squares of charged-lepton masses.  Note that the ratio $m_3/m_2
\alt 10$ for all $m_1$.  Naively, this argues against a simple see-saw
mass pattern and argues for the idea that the mass ratios can be
small, and quite possibly degenerate.

If the neutrino masses are degenerate, then the daughter neutrino
carries nearly the full energy of the parent neutrino in any reference 
frame (in the rest frame, a $\nu_2$ at rest decays to a $\nu_1$ nearly 
at rest).  This completely alters scenarios in which active daughters 
are detected.


\section{Neutrino Decay Models}

Non-radiative neutrino decay may arise through the coupling of the
neutrino to a very light or massless particle, such as a
Majoron~\cite{majoronmodels}.  Majoron models typically have
tree-level scalar or pseudoscalar couplings of the form
\begin{equation}
{\cal L}= g_{ij}\overline{\nu}_i \nu_j \chi + 
h_{ij} \overline{\nu}_i \gamma_5 \nu_j \chi + {\rm h.c.},
\label{coupl}
\end{equation}
where $\chi$ is a massless Majoron, which does not carry
definite lepton number.  For the couplings specified by
Eq.~(\ref{coupl}), the decay rates into neutrino and antineutrino
daughters are given by Ref.~\cite{kim}
\begin{eqnarray}
\Gamma_{\nu_2 \rightarrow \nu_1} 
&=& \frac{m_1 m_2}{16 \pi E_2}
\left[ g^2\left( \frac{x}{2} + 2 + \frac{2}{x}\ln x -\frac{2}{x^2}
-\frac{1}{2x^3} \right) \right. \, \nonumber \\
&+& \left. h^2\left( \frac{x}{2} - 2 + \frac{2}{x}\ln x +\frac{2}{x^2}
-\frac{1}{2x^3} \right) \right], \\
\Gamma_{\nu_2 \rightarrow \overline{\nu}_1} 
&=&  \frac{m_1 m_2}{16 \pi E_2} \left( g^2 + h^2 \right) 
\left[ \frac{x}{2} - \frac{2}{x}\ln x  -\frac{1}{2x^3} \right]\,,
\end{eqnarray}
where $x=m_2/m_1$, and we have dropped the subscripts on the coupling
constants.  The decay widths in this section are defined in the
laboratory frame, so the relation to the rest-frame lifetimes quoted
elsewhere is
\begin{equation}
\frac{\tau}{m} = \frac{1}{m\,\Gamma_{rest}} = \frac{1}{E\,\Gamma_{lab}}\,.
\end{equation}

In the limit of hierarchical neutrino masses, $m_2 \gg m_1$, the case
that has received the most attention to date, the decay rates are
equal:
\begin{equation}
\label{hier}
\Gamma_{\nu_2 \rightarrow \nu_1} = 
\Gamma_{\nu_2 \rightarrow \overline{\nu}_1} = 
\frac{(g^2+h^2)}{32\pi} \, \frac{m_2^2}{E_2} \,.
\end{equation}
In the opposite limit, where $m_2 \rightarrow m_1$ (but keeping
$\delta m^2 \neq 0$), we find instead:
\begin{eqnarray}
\Gamma_{\nu_2 \rightarrow \nu_1} 
&=& \frac{g^2}{16\pi} \, \frac{\delta m^2}{4 E_2} \; + \;
h^2 \times {\cal O}\left(\frac{\delta m^2}{m_2^2}\right)^3\,, \nonumber \\
\Gamma_{\nu_2 \rightarrow \overline{\nu}_1} 
&=& \frac{(g^2+h^2)}{16 \pi} \, \frac{m_2^2}{E_2}
\times {\cal O}\left(\frac{\delta m^2}{m_2^2}\right)^3\,.
\end{eqnarray}
The important point to note is that in these Majoron models, both
neutrino and antineutrino decay products may be produced, but the
relative weight of the two decay modes depends strongly upon the mass
hierarchy.

In the simplest versions of these models, the neutrino masses are
proportional to these coupling constants and hence the neutrinos are
exactly stable, as the matrix of coupling constants is diagonal in the
mass basis.  Even in this case, the neutrinos may have finite
lifetimes in matter, as the rotation of the mass basis in matter will
lead to non-diagonal couplings between matter mass
eigenstates~\cite{valle,matterdecay}.  In the most general case, the
basic models can easily be modified to permit non-diagonal couplings
in vacuum~\cite{fastdecay}.

While there are a huge variety of Majoron models in the literature
including, for instance, ``charged'' Majoron~\cite{burgess} and
``vector Majoron''~\cite{carone} models, we will not restrict our attention
to any particular model.  For example, instead of a Majoron model per
se, we can consider the couplings of neutrinos to a very light gauge
boson.  Similarly, we make no assumption about the relationship
between the neutrino masses and the couplings that give rise to decay,
or whether the neutrinos are Dirac or Majorana particles.

We thus take a purely phenomenological point of view and consider any
possible tree-level coupling between the neutrinos and a very light or
massless particle.  We shall consider the cases where the decay
products are active or sterile neutrinos or antineutrinos, plus an
``invisible'' particle.  While one could imagine models where such a
coupling is absent at tree level (as with radiative decay in the
Standard Model, for example), models of this sort are of less
interest, as any coupling arising only at loop level is likely to lead
to a very small decay rate.

Bounds on neutrino-Majoron couplings of the form in Eq.~(\ref{coupl})
may be obtained from considering their effects on the two-body
leptonic decays of $\pi$ and $K$ mesons at rest.  A nonzero coupling
allows the final neutrino to also appear as a neutrino or
antineutrino, plus a Majoron.  This increases the decay rate and also
smears the momentum distribution of the charged lepton.  The bounds
obtained are approximately $g^2 \alt 10^{-4}$~\cite{meson,britton}, 
and are reasonably
model-independent.\footnote{These bounds do not apply to the $g_{\tau
\tau}$ element of the coupling matrix which, given the large neutrino
mixing angles, will contribute to all elements of $g$ in the mass
basis.  Whether it is likely that $g_{\tau \tau}$ would be
significantly greater than all other (flavor basis) elements $g$ is a
model-dependent question of naturalness.}  
(Note that in this section we denote by $g$ either a scalar
or pseudoscalar coupling). A considerably more stringent bound may be
derived from limits on neutrinoless double beta decay with Majoron
emission~\cite{zuber}.  
However, the limit of $g^2 \alt 10^{-8}$ applies only to
the $g_{ee}$ element of the coupling matrix and does not directly
translate into a bound on the parameter of interest, namely $g_{21}$.

Translated into a bound on neutrino lifetimes, the meson-decay bound
on the coupling of $g^2 \alt 10^{-4}$ becomes 
\begin{equation}
\frac{\tau}{m} \agt
3 \times 10^{-5} \ \rm{s/eV}
\left(\frac{10^{-5} \ \rm{eV}^2}{\delta m^2}\right)\,,
\end{equation}
where we have used Eq.~(\ref{hier}).  With the solar $\delta m^2
\simeq 4 \times 10^{-5}$ eV$^2$, the lifetime limit obtained allows
substantial decay of solar neutrinos (from Table~I, the natural scale
of the problem is $\tau/m \simeq 10^{-4}$ s/eV).  If the mass
hierarchy is inverted, then solar neutrino decays can also occur with
$\delta m^2 \simeq 10^{-3}$ eV$^2$, in which case the derived lifetime
limit is 100 times weaker, and hence not useful.

We emphasize again that it is not our intention to restrict our
attention only to the case of Majoron models.  We discuss these models
only as a concrete example of the types of decay modes that may be
expected.  In Section~\ref{analysis} we take a quite general
perspective and discuss specific decay modes as illustrative,
model-independent examples.


\section{Solar $\nu_2$ Decay}
\label{analysis}

We consider the decay of neutrinos during their journey from the Sun
to Earth, and neglect decay inside the Sun, since the distances are
500 s and 2 s, respectively.  Decay rates can be increased in
matter~\cite{matterdecay} by the greater phase space provided by the
matter enhancement to $\delta m^2$.  However, for LMA-type solutions,
this effect is not large.  If decay in the Sun were significant it
would imply that our understanding of the solar neutrino flux is
fundamentally flawed; KamLAND and future solar-neutrino experiments
will check this possibility.

As noted, since an identifying characteristic of decay is its
particular energy dependence, we consider only the SK and SNO spectral
data.  These experiments are sensitive to $5-15$ MeV neutrinos from
$^8$B beta decay (while the normalization uncertainty is $\simeq 20\%$,
it is common to SK and SNO).
While one might think that the lowest energy
neutrinos are the best suited for testing decay effects, there are two
important caveats.  First, the gallium and chlorine radiochemical
experiments only measure energy-integrated rates, and receive
contributions from several different neutrino sources with uncertain
relative normalizations.  Exotic effects on the neutrino survival
probability can be hidden in this data.  Second, for the LMA
solutions, the $\nu_1$ to $\nu_2$ ratio rises at low energies, and as
noted, only the heavier $\nu_2$ can decay.

In the energy range covered by the SK and SNO spectra, the solar
neutrino flux is dominated by the $^8$B neutrinos. To a very good
approximation, these neutrinos are produced at the solar center where
the matter potential is
\begin{equation}
\zeta \equiv {\sqrt 2} G_F N_e \frac{2 E}{\delta m^2} 
= 15.3 \left( \frac{E}{10\ {\rm MeV}} \right) 
\left( \frac{10^{-5}{\ \rm eV^2}}{\delta m^2} \right).
\end{equation}
The initial $\nu_1$ and $\nu_2$ fractions of the flux are specified by
\begin{eqnarray}
P_{\nu_1}^i &=& \cos^2 \theta_m, \nonumber \\
P_{\nu_2}^i &=& \sin^2 \theta_m,
\end{eqnarray}
where the matter mixing angle given by
\begin{equation}
\sin^2{2 \theta_m} = 
\frac{\sin^2{2 \theta_v}}{\sin^2{2 \theta_v}+(\cos{2\theta_m}-\zeta)^2}\,,
\end{equation}
and $\theta_v$ is the vacuum angle.  As representative values in the
LMA allowed region~\cite{lma}, we choose $\delta m^2 = 4 \times
10^{-5}$ eV$^2$ and $\sin^2{2\theta_v} = 0.9$.  The MSW transformation
is always adiabatic in the solar neutrino energy range for the LMA
solutions, so the fluxes of $\nu_1$ and $\nu_2$ are unchanged at the
solar surface (neglecting the tiny decay probability over the solar
radius).  The final $\nu_1$ and $\nu_2$ fluxes at Earth, taking decay 
into account are
\begin{eqnarray}
\label{finalsterile}
P_{\nu_1}^f &=& P_{\nu_1}^i\,, \nonumber \\
P_{\nu_2}^f &=& P_{\nu_2}^i 
\exp\left(-\frac{L}{E} \times \frac{m}{\tau}\right)\,.
\end{eqnarray}
Note that the small variation in the Earth-Sun distance cannot be
exploited, as its effects are washed out by the energy resolution.
It is convenient to think of a typical energy for decay (that makes
the argument of the exponential unity); in convenient units, this is
\begin{equation}
\frac{E_{decay}}{1 {\rm\ MeV}} = 
\frac{5 \times 10^{-4} {\rm\ s/eV}}{\tau/m}\,.
\label{edecay}
\end{equation}
The electron neutrino survival probability at Earth is thus
\begin{equation}
P(\nu_e \rightarrow \nu_e) = 
P_{\nu_1}^f \cos^2\theta_v + P_{\nu_2}^f \sin^2\theta_v\,.
\end{equation}
The muon neutrino appearance probability is
\begin{equation}
P(\nu_e \rightarrow \nu_\mu) = 
P_{\nu_1}^f \sin^2\theta_v + P_{\nu_2}^f \cos^2\theta_v\,.
\end{equation}
Because of decay, $P(\nu_e \rightarrow \nu_e) + P(\nu_e \rightarrow
\nu_\mu) < 1$ (at this point, we are considering all decay products
to be sterile).

In the absence of decay or oscillations, the recoil electron spectrum
from neutrino-electron scattering in SK can be computed by convolving
the $^8$B neutrino spectrum~\cite{ortiz} with the $\nu_e - e$ elastic
scattering cross section $d\sigma/dT$ (see, e.g.,
Ref~\cite{magnetic}), and smearing with the SK energy
resolution~\cite{SKres}.  For a given decay or oscillation scenario,
the spectrum is calculated in the same way, but now allowing an
energy-dependent survival probability for the neutrinos.  The
contribution (about 6 times smaller, and with a slightly different
differential cross section) from $\nu_\mu - e$ elastic scattering must
also be included.  These two spectra, binned every 0.5 MeV in the
recoil electron total energy $E_e = T_e + m_e$ can then be divided
bin-by-bin.  The resulting ``ratio spectrum'' $R_{SK}$ closely
approximates how SK presents their data, though they of course use a
much more detailed Monte Carlo to model the detector response.  Also,
in their case the numerator in the ratio is the measured counts per
bin.


\subsection{Simple $\nu_2$ Disappearance}
\label{simple}

\begin{figure}
\begin{center}
\epsfig{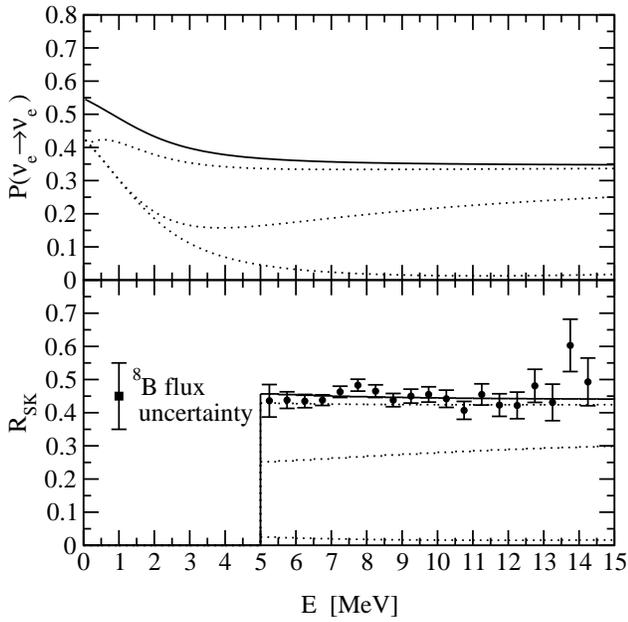}
\caption{\label{basic} In the upper panel, the electron neutrino
survival probability is shown versus {\it neutrino} energy.  In the
lower panel, the ratio of measured to expected spectra in SK is shown
versus the recoil {\it electron} total energy, as detected in
neutrino-electron scattering above 5 MeV.  The solid lines correspond
to the LMA solution ($\delta m^2 = 4 \times 10^{-5}$ eV$^2$,
$\sin^2{2\theta} = 0.9$) with stable $\nu_2$.  The dashed lines, in
order of decreasing height, correspond to $\nu_2$ lifetimes of $\tau/m
= 10^{-3}, 10^{-4},{\rm and\ } 10^{-5}$ s/eV (by Eq.~(\ref{edecay}),
these correspond to decays at typical energies of 0.5, 5, and 50 MeV,
respectively).
The 1258-day data from
SK are shown in the lower panel, as is the size of the flux
normalization uncertainty.  The decay products of $\nu_2$ are
considered to be sterile.}
\end{center}
\end{figure}

In the upper panel of Fig.~\ref{basic} we display the $\nu_e$ survival
probability as a function of neutrino energy, for a range of decay
rates.  We have chosen parameters in the LMA allowed region.  
The solid curve corresponds to the case of no decay.
In the lower panel, we show the corresponding SK
electron energy spectrum, which also includes the $\nu_\mu$ component.
At large energies, the height of the solid lines is set just by
$\sin^2{2\theta} = 0.9$; SNO expects 0.34 and SK expects $0.34 +
0.66/6 = 0.45$.  The upturn at low energies is set by $\delta m^2 = 4
\times 10^{-5}$ eV$^2$.  Note that the chlorine experiment is mostly
sensitive to the $^8$B neutrinos, so also expects $\simeq 1/3$.  In
order to correctly calculate the rate in the gallium detectors, one
must also integrate over the finite source region in the Sun.
One can see from the figure that as the neutrino lifetime
approaches $10^{-4} {\rm\ s/eV}$ or shorter the spectral shape begins
to display a significant deviation from the flat SK ratio spectrum.
The total flux also begins to depart significantly from the measured
value, though only at about the $2\sigma$ level, taking into account
the $^8$B flux uncertainty.

\subsection{Compensating Effects from Oscillations}

In Fig.~\ref{basic} we used the flatness of the SK spectrum to
estimate a bound on the decay rate of $\nu_2$.  However,
things may not be as straightforward as this basic case.  In
particular, since oscillation effects may lead to an
energy distortion, we must address the possibility that a cancellation
between oscillation and decay survival probabilities may produce an
acceptably flat spectrum.

For example, we could choose a $\delta m^2$ somewhat larger than those
in the allowed LMA region, so that the $\nu_e$ survival probability
begins to rise toward the lower energy end of the SK energy range.
Since the decay rate is larger for lower energy neutrinos,
the question then becomes whether the decay might be ``just so'' in
the sense that it conspires to cancel this tilt in the survival
probability, resulting in a spectrum which is apparently flat.  Such
a scenario has recently been addressed in
Ref.~\cite{choubey,chou,joshipura}.  We view this scenario as somewhat
unnatural, however we agree that it cannot be excluded.

\begin{figure}
\begin{center}
\epsfig{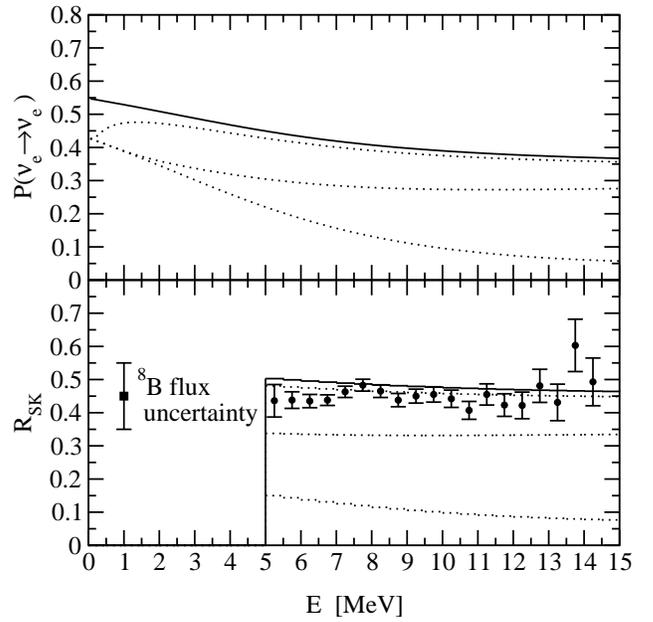}
\caption{\label{conspire} An example of the parameter degeneracy
between the lifetime and the mixing parameters is illustrated.  The
description is as for Fig.~\ref{basic}, except that now $\delta m^2 =
12 \times 10^{-5}$ eV$^2$, which flattens $R_{SK}$ for $\tau/m =
10^{-4}$ s/eV, making it consistent with the SK spectral shape.  It is
about $1\sigma$ discrepant in terms of normalization.}
\end{center}
\end{figure}

In Fig.~\ref{conspire}, we show an example of such a conspiracy, using
oscillation parameters for which, in the absence of decay, the SK
spectrum is not very flat.  Here, if we choose a lifetime of order
$10^{-4} {\rm\ s/eV}$ we see that a reasonably flat spectrum may be
regained.  Note however, the large difference between the solid and
dashed curves in this figure, both in terms of the spectral shape and 
the overall
rates. It is important to realize that KamLAND will essentially break
the degeneracy between the oscillation parameters and the neutrino
lifetime, giving a {\it prediction} for the mixing effects (without
decay) on the solar neutrino spectrum.  One can then go and look for a
deviation from this prediction in the solar neutrino spectrum.  As is
clear from Figs.~\ref{basic} and \ref{conspire}, a lifetime of order
$10^{-4} {\rm\ s/eV}$ defines the scale at which decay could possibly
be distinguished.

\subsection{Appearance of Active Daughter $\nu_1$}
\label{degendaughters}

Another possible complication
is the possible detection of neutrinos produced as
decay products.\footnote{The possible detection of decay products was
actually noted in the first paper on neutrino decay,
Ref.~\cite{solardecay}.  However, active daughter neutrinos (as opposed to 
antineutrinos) have since received very little attention in the literature.} 
The replacement of a parent neutrino with an active daughter may partially
hide the effects of decay.  Unlike the situation discussed above, this
would not be sensitive to a conspiracy between parameters.  Rather,
the appearance and possible detection of active decay products is a
quite generic expectation, as most plausible decay models will feature
a neutrino or antineutrino in the final state.  There are a range of
model-dependent possibilities for these final-state neutrinos, such
as whether the neutrinos are Dirac or Majorana particles and whether
the decay products are active or sterile~\cite{dodelson}.
In this subsection we study the effects of active daughter neutrinos.

If active daughters are produced, the overall neutrino mass scale is
important.  In virtually all studies of neutrino decay, it has been
assumed that the neutrino mass spectrum is hierarchical, and thus the
mass of the final state neutrino may be neglected.  This is important,
since the energy difference between initial and final state neutrinos
depends on the values of the masses.  If a neutrino decays to a
massless particle and another neutrino, the fraction of the initial
neutrino energy carried by the final state neutrino ranges from a
maximum of $E_f/E_i \simeq 1$ to a minimum of $E_f/E_i \simeq
m_1^2/m_2^2$.  Where a hierarchical mass spectrum is a good
approximation, the two decay products will each carry a large fraction
of the original neutrino energy, hence the daughter neutrino will, on
average, be degraded in energy.  However, if the neutrino masses are
degenerate, the daughter neutrino must have approximately the same
energy as the parent neutrino.  For example, if we take $\delta m^2
\simeq 4 \times 10^{-5}\rm{\ eV}^2$ and the degenerate mass larger 
than about
$m_1\simeq m_2 \simeq 0.03 {\rm\ eV}$ the energy degradation of the
daughter neutrino would be less than about 0.5 MeV -- the width of the
SK energy bins.

Given this, one may wonder if it is possible to set any decay bound at
all.  For example, if the mixing angle were exactly maximal, any decay
of $\nu_2$ to $\nu_1$ in which the neutrino energy was not degraded
would be undetectable.  However, the solar $\nu_e$ survival
probability appears to be less than 1/2 in the SNO/SK and Cl energy
region (and somewhat larger in the Ga region) strongly disfavoring 
exact maximal mixing.  As
long as the mixing angle is not exactly maximal, decay will cause the
relative fluxes of $\nu_e$ and $\nu_{\mu}$ to vary as a function of
energy, which would be distinguishable in both SK and
SNO.

\begin{figure}
\begin{center}
\epsfig{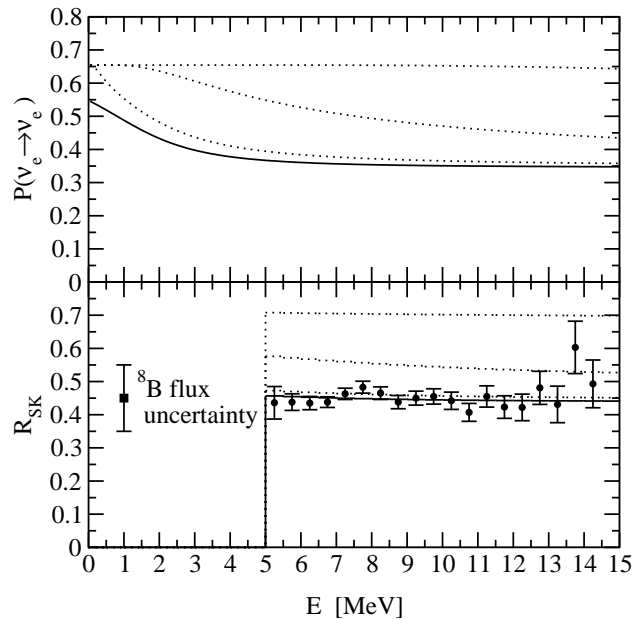}
\caption{\label{daughters} The effect of $\nu_2$ decay to active $\nu_1$ 
daughters is
shown.  We have assumed $m_1$ and $m_2$ are nearly degenerate, so
that the daughter energy is approximately the full energy of the
parent. The description is as for Fig.~\ref{basic}, except that now
the dotted lines, in order of {\it increasing} height, correspond to
$\nu_2$ lifetimes of $\tau/m = 10^{-3}, 10^{-4},{\rm and\ } 10^{-5}$
s/eV.  Again, the case $\tau/m = 10^{-4}$ s/eV is discrepant in terms of
shape and normalization.}
\end{center}
\end{figure}

In Fig.~\ref{daughters} we demonstrate the effect on the spectrum if
$\nu_2$ decays to the orthogonal state $\nu_1$, in the limit where the
daughter neutrinos carry the full energy of the parent.\footnote{The 
possibility that the entire flux (across the whole
energy range) has decayed to the $\nu_1$ state by the time the
neutrinos reach Earth would give an energy-independent
suppression of the solar $\nu_e$ flux, contrary to that observed.}
The expressions in Eq.~(\ref{finalsterile}) have now to be replaced by
\begin{eqnarray}
\label{finaldaughters}
P_{\nu_1}^f &=& 
P_{\nu_1}^i + P_{\nu_2}^i 
\left[1 - \exp\left(-\frac{L}{E} \times \frac{m}{\tau}\right) \right], 
\nonumber \\
P_{\nu_2}^f &=& 
P_{\nu_2}^i \exp\left(-\frac{L}{E} \times \frac{m}{\tau}\right).
\end{eqnarray}
in order to include the $\nu_1$ produced by the decay of $\nu_2$.
Note that there is no quantum-mechanical interference between $\nu_1$
originally in the beam with those produced by the decays of 
$\nu_2$~\cite{ohlsson}.  In addition, we may take the daughter
neutrinos to be collinear with the parent neutrinos~\cite{ohlsson}.

An interesting feature of Fig.~\ref{daughters} is the direction of the
deviation from a flat spectrum.  Rather than a deviation downward (as
observed in Fig.~\ref{basic}) caused by the depletion of the flux at
lower energies, we instead obtain a deviation upward.  This is due to
the replacement of a portion of the $\nu_2$ flux by $\nu_1$, which has
a larger $\nu_e$ component (since $\theta_{solar} \simeq 36^\circ$).  
Note, however, that the size of $\tau/m$
where decay effects show up as a significant deviation (either to the
flat spectrum, or to the total rates) is comparable to that shown in
Fig.~\ref{basic}.  Therefore, the lifetime limits for the cases of
decay to steriles and decay to active neutrinos of non-degraded energy
could be comparable (if $m_1 \simeq m_2$ were known).

\subsection{Dark-Side Inversion Undone by Decay}

\begin{figure}
\begin{center}
\epsfig{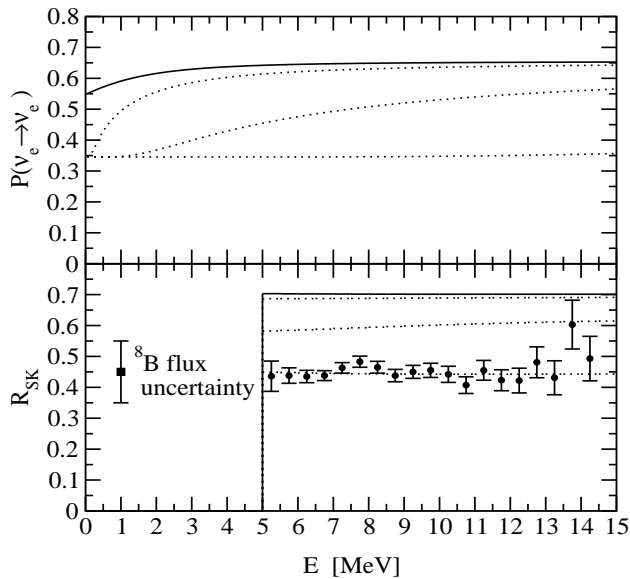}
\caption{\label{darkside} The effect of $\nu_2$ decay to active
$\nu_1$ daughters is shown again, this time with a ``dark-side'' angle
of $54^\circ$ instead of $36^\circ$ (but still with $\delta m^2 = 4
\times 10^{-5}$ eV$^2$, $\sin^2{2\theta} = 0.9$).  The description is
as for Fig.~\ref{basic}, except that now the dotted lines, in order of
{\it decreasing} height, correspond to $\nu_2$ lifetimes of $\tau/m =
10^{-3}, 10^{-4},{\rm and\ } 10^{-5}$ s/eV.  In this unusual case, a
lifetime {\it as short as} $10^{-5}$ s/eV is required to undo the
$\nu_2/\nu_1$ reversal in $\nu_e$ caused by choosing a ``dark-side'' 
angle.}
\end{center}
\end{figure}

We now consider an unusual case that requires decay.  Suppose the 
solar neutrino
parameters live on the so-called ``dark-side''~\cite{dark} of the
parameter space, that is, the hierarchy is inverted such a way that
$\nu_2$ has a larger $\nu_e$ component than does $\nu_1$.  This
situation does not provide a good fit to the solar neutrino data, as
an MSW resonance would not take place in the Sun and it is not
possible to obtain a $\nu_e$ survival probability less than one
half.  However, if we add fast decay to this dark-side solution, we can
convert the entire flux to $\nu_1$, obtaining a solution that is in
good agreement with the SK, SNO and Cl data.  That is, decay
effectively undoes the $\nu_2 - \nu_1$ reversal in $\nu_e$.  
We have plotted such
an example in Fig.~\ref{darkside}.  Note that the curve which most
closely resembles the measured SK data is the one which corresponds to
the largest decay rate.  The only problem with this solution lies with
Ga flux measurement of greater than a half -- a difficulty which,
alone, is probably sufficient to rule out such a solution.  So,
although such a scenario seems unlikely, it is enough, however, to
give one pause for thought.

\subsection{Other Possibilities}

In Fig.~\ref{basic} we have considered decay products that are
completely undetectable (this could also be achieved if the daughter
neutrino energies were artificially taken to zero), 
while in Fig.~\ref{daughters} we considered
active daughters of the full energy, such that their detectability is
optimized.  It is clear that the case where the daughter neutrino
should be active but have degraded energy, lies somewhere between the
examples considered in Fig.~\ref{basic} and Fig.~\ref{daughters}.
This case would occur if $m_1 \ll m_2$.
However, note that this directly modifies the absolute spectrum, which
is steeply falling (see Ref~\cite{SK}); the ratio spectra shown above
cannot be averaged by eye.
By comparing Figs.~\ref{basic} and \ref{daughters}, it should be clear
that it could be quite difficult to discern such a decay using either
the spectral shape or the total flux

Another example is that of the inverted hierarchy, where the solar
neutrino flux can decay to the eigenstate that consists mainly of
$\nu_{\mu}$ and $\nu_{\tau}$.  Since these flavors have a
cross-section approximately 6 times smaller than $\nu_e$, this
situation is closer to the sterile daughters example presented in
Fig.~\ref{basic} than to the active daughter $\nu_1$ case presented in
Fig.~\ref{daughters}.

The decay products might include antineutrinos rather than neutrinos
or, more generally, a mixture of the two.  Since
$\overline{\nu}_{\mu}$ and $\overline{\nu}_{\tau}$ have similar
neutrino-electron scattering cross-sections to $\nu_{\mu}$ and
$\nu_{\tau}$, they would also be difficult to detect.

In Ref.~\cite{pakvasadaughters,choubey}, the appearance of 
$\bar{\nu}_e$ as a decay
product is discussed.  An estimated limit on the solar
$\bar{\nu}_e$ flux has been obtained for the SK data in
Ref.~\cite{torrente}, using the technique of Ref.~\cite{invbeta}, and
is quoted as 3.5\% of the $^8$B $\nu_e$ flux (at the 95\% C.L.).
Ref.~\cite{choubey} takes this to
imply a stringent decay limit of $\tau/m \simeq 10^{-3}$ s/eV.  Though
they state that the $\bar{\nu}_1$ carries less energy than the parent
neutrino, it appears that the bound is derived without taking this
into account.  Indeed, as we have stressed above, the daughter may
carry nearly the full energy of the parent.  But if not, it makes a
big difference for the yield, since the $\bar{\nu}_e p \rightarrow e^+
n$ cross section is nearly quadratic in neutrino energy.  And the
neutrino spectrum is falling steeply with energy, so decay products
from a high energy can be hidden at a lower energy.  Also, if the
daughter energy is less than the parent energy, then the limit as
quoted from Ref.~\cite{torrente} does not apply.  The limit of 3.5\%
assumes that the $\bar{\nu}_e$ spectrum has the same shape as
the $^8$B $\nu_e$ spectrum.  If that assumption is relaxed, then the
flux limit is more conservatively about 10\%.  Thus the decay limit is
certainly considerably weaker than $\tau/m \simeq 10^{-3}$ s/eV. 

Finally, one might reasonably ask if the solar neutrino data can be
{\it explained} by neutrino decay.  The formulation of this question
that seems to be the most interesting is this: Can the solar $\nu_e$
and LSND \cite{lsnd} $\bar{\nu}_e$ data be made consistent?  
The combined solar,
atmospheric, and LSND results require three independent $\delta m^2$
values, whereas only two are allowed in three-generation mixing.
Thus, our question can be rephrased: Can one explain all of the data
with the two $\delta m^2$ values and one $\tau/m$ value?  Let us
assume the LSND vacuum mixing parameters.  For the large $\delta m^2
\simeq 1$ eV$^2$, matter effects in the Sun are negligible.  We are
free to invert the sign of $\delta m^2$ so that the solar $\nu_e
\simeq \nu_2$ (the LSND mixing angle is very small).  Suppose decay
turns this into $\simeq 1/3\,\nu_2 + 2/3\,\nu_1 \simeq 1/3\,\nu_e +
2/3\,\nu_\mu$ in the SK energy region, roughly matching the SNO and SK
observations (we ignore the spectral distortion).  But then decay is
complete in the gallium-detector energy range, predicting nearly zero
$\nu_e$ flux there, in gross disagreement with observations.


\section{Other Neutrino Decays}

Let us suppose that a model-independent limit on solar neutrino decay
can be arrived at, and that it reaches the scale of $\tau/m \alt
10^{-4}$ s/eV.  Once such a limit has been established, can it be used
to set meaningful limits on the possible decay of neutrinos from other
sources?  In particular, we shall consider the decay of atmospheric
neutrinos.  It is immediately obvious that if we restrict ourselves to
the three active neutrinos, it is difficult to arrange for decay to
play a role in the atmospheric neutrino problem.  Since two of the
three mass eigenstates have large $\nu_e$ components, it is hard to
see how decay could alter the atmospheric $\nu_{\mu}$ flux, while
leaving the $\nu_e$ flux unaltered, as the data suggests.

We can make this point more precise.  Let us define the neutrino mass
eigenstates to be such that $m_3 > m_2 > m_1$.  Considered as a
function of mass squared, in a normal hierarchy two states are close
together (the solar $\delta m^2$), and are well below the third state
(by the atmospheric $\delta m^2$).  The sign of the solar $\delta m^2$
is fixed by the observation of matter effects in the Sun (for the
opposite sign, there is no MSW resonance).  However, the sign of the
atmospheric $\delta m^2$ that dictates vacuum oscillations is
unknown.  An inverted hierarchy is obtained when this state is well
{\it below} the solar doublet on a mass-squared scale.  See also
Fig.~\ref{levels}, but note that the scale there is mass, not mass
squared.

If we have a normal hierarchy, the solar decay limit applies to
$\tau_{21}/m_2$, whereas if we have an inverted hierarchy, the limit
applies to both $\tau_{32}/m_3$ and $\tau_{31}/m_3$.  In the inverted
case, only $\nu_2 \rightarrow \nu_1$ is not directly constrained (in
the inverted case, we relabel so that $m_3 > m_2 > m_1$).  However,
since this decay mode has virtually the same phase space as $\nu_3
\rightarrow \nu_1$, it must also be constrained, unless there were a
large hierarchy in the respective couplings.  It is important to
remember that the flavor content of the mass eigenstates is irrelevant
for their decays.

In the case of a normal hierarchy, we can translate the bound on
$\nu_2$ decay into one on $\nu_3$ if we make certain assumptions.
Since the quantity limited by experiment is $\tau/m_2$ and $\tau/m_2
\sim (g^2 \delta m^2)^{-1}$ (using Eq.~(\ref{hier})), then the overall
mass scale is mostly irrelevant (at least for the hierarchical case)
for setting a bound on $g$.  A solar decay limit of $10^{-4}$ s/eV
would translate into a limit on the coupling of
\begin{equation}
g^2_{21} \alt 3 \times 10^{-5} 
\left(\frac{10^{-5}{\rm\ eV}^2}{\delta m^2}\right)\,.
\end{equation}
For the solar LMA solution, $\delta m^2 \simeq 4 \times 10^{-5}$
eV$^2$, one would obtain a limit of $g^2_{21} < 10^{-5}$,
slightly more restrictive than the limit from meson decays of 
$g^2 \alt 10^{-4}$.  However, from Section~II we see that the
correspondence between neutrino lifetime and the coupling constant
depends sensitively on the mass hierarchy.

In order to translate a bound on $\nu_2$ into one on $\nu_3$, we would
have to assume that $g$ is approximately universal among generations.
Of course, this is a very model-dependent assumption.  The limit on
the $\nu_3$ lifetime would become
\begin{equation}
\frac{\tau_{31}}{m_3} \sim \frac{\tau_{21}}{m_2} \times
\frac{\delta m^2_{21}}{\delta m^2_{31}}
\agt 10^{-6} {\rm\ s/eV}\,.
\end{equation}
The natural scale for setting a limit on neutrino decay with
atmospheric neutrinos is $\tau/m \simeq L/E \simeq 10^4 {\rm\ km}/ 300
{\rm\ MeV} \simeq 10^{-10}$ s/eV.  Thus if there were significant decay
in the atmospheric neutrinos, a significant hierarchy in the couplings
$(g_{31}/g_{21})^2 \sim 10^4$ would be required.  The exception to
this is to greatly increase the $\delta m^2$, as in
Ref.~\cite{barger}.  This model is now strongly disfavored since it
cannot accommodate large-angle mixing with an active neutrino in the
solar sector.  Finally, as noted, if there is an inverted hierarchy, a
limit on solar neutrino decay may apply directly to atmospheric
neutrino decay, since the same states may be involved.

A limit on solar neutrino decay would also apply to decays to
``phantom'' neutrinos.  Light sterile neutrinos can be used to add new
mass eigenstates anywhere relative to the standard hierarchy.  These
new mass eigenstates might be {\it inaccessible by flavor mixing} (or
with very small angles), but might be reached via decays (which
connect mass eigenstates directly). If so, the presence of these
phantom neutrinos could result in an {\it apparent} non-unitarity of
the $3 \times 3$ mixing matrix.  Indeed, such tests could infer their
existence even if decay were not seen directly.
These phantom neutrinos can also be lower in mass than $\nu_1$ in the 
standard case; however, the very long SN1987a lifetime limit~\cite{frieman} 
would apply to those states.


\section{Conclusions}

Solar and supernova neutrinos have been observed on Earth suggesting,
naively, that they do not decay.  However, this conclusion cannot
immediately be drawn unless one can rule out certain subtleties.  When
considering these possibilities we must be careful to distinguish the
mass eigenstates (for which the lifetimes are properly defined) and
the flavor eigenstates.

\begin{enumerate}

\item
While the LMA solution is an excellent fit to the solar neutrino data,
one cannot completely rule out more exotic scenarios until
KamLAND~\cite{kamland} confirms the LMA parameters in an experiment
that does not rely upon properties of the Sun.  KamLAND will determine
the ``solar'' mixing parameters using antineutrinos rather than
neutrinos, in vacuum rather than in matter, and using a much shorter
baseline than the Earth-Sun distance.  In addition, exotic effects
such as flavor-changing neutral currents, or resonant spin-flip
transitions can be eliminated.  Only then shall we have the complete
confidence in the LMA solution necessary to fully utilize the
potential of the solar neutrino beam as a probe of non-standard
neutrino properties.

\item
The flat energy spectrum observed in SK and SNO is well described by
the LMA solution, which would seem to argue against an energy-dependent 
distortion of the survival probability, as would be
characteristic of decay.  However, for parameters outside the LMA
region one can obtain an energy dependent oscillation survival
probability to offset the effects of decay.  This possibility may be
eliminated by KamLAND.

\item
Decay will typically (though not always) cause a depletion of the
total solar neutrino flux.  While this is also a possible way to
identify decay, we should keep in mind the $20\%$ uncertainly in the
$^8$B flux normalization.  That uncertainty  will be reduced by
future neutral-current data from SNO, as well as direct nuclear-physics
measurements~\cite{s17}.

\item
In the limit that the neutrino masses are degenerate, a daughter
neutrino produced by decay will carry the full energy of the parent
neutrino, and could be detected in solar neutrino experiments.  The
replacement of the parent neutrino with an active daughter of the same
energy could obscure the characteristic features of decay.  This is
especially pertinent in the case of the decay $\nu_2 \rightarrow
\nu_1$, where both $\nu_2$ and $\nu_1$ have large $\nu_e$ projections.
 
\item
Decays producing $\overline{\nu}_e$ daughters should be readily
detectable, provided the energy of the $\overline{\nu}_e$ is not too
degraded.  If the hierarchy is inverted, there may be decays to the
state that is dominantly $\nu_\mu$ and $\nu_\tau$ (or the antiparticle
state).  These are harder to detect in SK, though they would add to
the integral neutral-current rate in SNO.

\item
If a model-independent limit on neutrino decay can be established, the
bound will likely be of order $\tau/m \agt 10^{-4}$ s/eV.  This is
close to the limit obtained via meson-decay bounds (which may or
may not apply; see above).  Though the bound is quite weak, the solar
neutrino flux would, however, still provide the best limit on
non-radiative neutrino decay.  This limit is about eight orders of
magnitude too weak to rule out the decay of astrophysical neutrinos
(e.g., from a Galactic supernova or a distant AGN \cite{Keranen}) on 
their journey to
Earth.  We thus have absolutely no guarantee we shall be able to use
such neutrinos to probe the astrophysics of these sources without
taking decay into account.

\end{enumerate}


\medskip
\section*{Acknowledgments}

We thank Kev Abazajian, Boris Kayser, Gail McLaughlin, Stephen Parke,
David Rainwater, and Mark Vagins for useful discussions.  We also
thank B. Peon for pointing out the existence (or non-existence) of
Hinchliffe's Rule~\cite{peon}.  J.F.B. (as the David N. Schramm
Fellow), and N.F.B. were supported by Fermilab, which is operated by
URA under DOE contract No.\ DE-AC02-76CH03000, and were additionally
supported by NASA under NAG5-10842.



\begin{thebibliography}{99}

\bibitem{magnetic}
J.~F.~Beacom and P.~Vogel, Phys.\ Rev.\ Lett.\  {\bf 83}, 5222 (1999).

\bibitem{SK} 
S.~Fukuda {\it et al.}, Phys.\ Rev.\ Lett.\  {\bf 86}, 5651 (2001);
S.~Fukuda {\it et al.}, Phys.\ Rev.\ Lett.\  {\bf 86}, 5656 (2001).

\bibitem{SNO} 
Q.~R.~Ahmad {\it et al.}, Phys.\ Rev.\ Lett.\  {\bf 87}, 071301 (2001).

\bibitem{solardecay}  
J.~N.~Bahcall, N.~Cabibbo and A.~Yahil, 
Phys.\ Rev.\ Lett.\  {\bf 28}, 316 (1972).

\bibitem{solardecay1} 
S.~Pakvasa and K.~Tennakone, Phys.\ Rev.\ Lett.\  {\bf 28}, 1415 (1972).

\bibitem{raffelt}
G.G. Raffelt, {\it Stars as Laboratories for Fundamental Physics}
(University of Chicago Press, Chicago, 1996).

\bibitem{RPP}
D.~E.~Groom {\it et al.}, Eur.\ Phys.\ J.\ C {\bf 15}, 1 (2000).

\bibitem{cosmolimits}
M.~Kaplinghat, R.~E.~Lopez, S.~Dodelson and R.~J.~Scherrer,
Phys.\ Rev.\ D {\bf 60}, 123508 (1999);
S.~Hannestad, Phys.\ Rev.\ Lett.\  {\bf 85}, 4203 (2000);
S.~Hannestad, Phys.\ Lett.\ B {\bf 431}, 363 (1998);
M.~J.~White, G.~Gelmini and J.~Silk, Phys.\ Rev.\ D {\bf 51}, 2669 (1995).

\bibitem{tritium}
Ch. Weinheimer et al., Phys. Lett. {\bf B460}, 219 (1999); \\
V.M. Lobashev et al.,  Phys. Lett. {\bf B460}, 227 (1999); \\
J.~Bonn {\it et al.}, Nucl.\ Phys.\ Proc.\ Suppl.\  {\bf 91}, 273 (2001).

\bibitem{majoronmodels}
Y.~Chikashige, R.~N.~Mohapatra and R.~D.~Peccei,
Phys.\ Lett.\ B {\bf 98}, 265 (1981);
G.~B.~Gelmini and M.~Roncadelli, Phys.\ Lett.\ B {\bf 99}, 411 (1981).

\bibitem{kim}
C.~W.~Kim and W.~P.~Lam, Mod.\ Phys.\ Lett.\ A {\bf 5}, 297 (1990);
C.~Giunti, C.~W.~Kim, U.~W.~Lee and W.~P.~Lam,
Phys.\ Rev.\ D {\bf 45}, 1557 (1992);
Z.~G.~Berezhiani, G.~Fiorentini, M.~Moretti and A.~Rossi,
Z.\ Phys.\ C {\bf 54}, 581 (1992).

\bibitem{valle}
M.~Kachelriess, R.~Tomas and J.~W.~Valle, 
Phys.\ Rev.\ D {\bf 62}, 023004 (2000).

\bibitem{matterdecay} 
Z.~G.~Berezhiani and M.~I.~Vysotsky, Phys.\ Lett.\ B {\bf 199}, 281 (1987).

\bibitem{fastdecay}
G.~B.~Gelmini and J.~W.~Valle, Phys.\ Lett.\ B {\bf 142}, 181 (1984);
A.~Acker, A.~Joshipura and S.~Pakvasa, Phys.\ Lett.\ B {\bf 285}, 371 (1992).

\bibitem{burgess}
C.~P.~Burgess and J.~M.~Cline, Phys.\ Lett.\ B {\bf 298}, 141 (1993).

\bibitem{carone}
C.~D.~Carone, Phys.\ Lett.\ B {\bf 308}, 85 (1993).

\bibitem{meson}
V.~D.~Barger, W.~Y.~Keung and S.~Pakvasa, Phys.\ Rev.\ D {\bf 25}, 907 (1982).

\bibitem{britton}
D.~I.~Britton {\it et al.}, Phys.\ Rev.\ D {\bf 49}, 28 (1994);
C.~E.~Picciotto {\it et al.}, Phys.\ Rev.\ D {\bf 37}, 1131 (1988).

\bibitem{zuber}
K.~Zuber, Phys.\ Rept.\  {\bf 305}, 295 (1998).

\bibitem{lma}
G.~L.~Fogli, E.~Lisi, D.~Montanino and A.~Palazzo,
Phys.\ Rev.\ D {\bf 64}, 093007 (2001);
J.~N.~Bahcall, M.~C.~Gonzalez-Garcia and C.~Pena-Garay,
JHEP {\bf 0108}, 014 (2001);
A.~Bandyopadhyay, S.~Choubey, S.~Goswami and K.~Kar,
Phys.\ Lett.\ B {\bf 519}, 83 (2001).

\bibitem{ortiz}
C.~E.~Ortiz, A.~Garcia, R.~A.~Waltz, M.~Bhattacharya and A.~K.~Komives,
Phys.\ Rev.\ Lett.\  {\bf 85}, 2909 (2000).

\bibitem{SKres}
M.~Nakahata {\it et al.}, Nucl. Instrum. Meth. {\bf A421}, 113 (1999).

\bibitem{chou} 
C.~K.~Chou and M.~Chou, Phys.\ Scripta {\bf 64}, 197 (2001).

\bibitem{choubey} 
A.~Bandyopadhyay, S.~Choubey and S.~Goswami,
Phys.\ Rev.\ D {\bf 63}, 113019 (2001).

\bibitem{joshipura}
A.~S.~Joshipura, E.~Masso and S.~Mohanty, hep-ph/0203181.

\bibitem{dodelson}
S.~Dodelson, J.~A.~Frieman and M.~S.~Turner,
Phys.\ Rev.\ Lett.\  {\bf 68}, 2572 (1992).

\bibitem{ohlsson}
M.~Lindner, T.~Ohlsson and W.~Winter, Nucl.\ Phys.\ B {\bf 607}, 326 (2001).

\bibitem{dark}
A.~de Gouvea, A.~Friedland and H.~Murayama,
Phys.\ Lett.\ B {\bf 490}, 125 (2000);
G.~L.~Fogli, E.~Lisi and D.~Montanino, Phys.\ Rev.\ D {\bf 54}, 2048 (1996).

\bibitem{pakvasadaughters}
R.~S.~Raghavan, X.~G.~He and S.~Pakvasa,
Phys.\ Rev.\ D {\bf 38}, 1317 (1988).

\bibitem{torrente}
E.~Torrente-Lujan, Phys.\ Lett.\ B {\bf 494}, 255 (2000).

\bibitem{invbeta}
P.~Vogel and J.~F.~Beacom, Phys.\ Rev.\ D {\bf 60}, 053003 (1999).

\bibitem{lsnd}
A.~Aguilar {\it et al.}, Phys.\ Rev.\ D {\bf 64}, 112007 (2001).

\bibitem{barger}
V.~D.~Barger {\it et al.}, Phys.\ Lett.\ B {\bf 462}, 109 (1999);
V.~D.~Barger {\it et al.} Phys.\ Rev.\ Lett.\  {\bf 82}, 2640 (1999);
G.~L.~Fogli {\it et al.}, Phys.\ Rev.\ D {\bf 59}, 117303 (1999).

\bibitem{frieman}
J.~A.~Frieman, H.~E.~Haber and K.~Freese,
Phys.\ Lett.\ B {\bf 200}, 115 (1988).

\bibitem{kamland} 
A.~Piepke, Nucl.\ Phys.\ Proc.\ Suppl.\  {\bf 91}, 99 (2001).

\bibitem{s17}
A.~R.~Junghans {\it et al.}, Phys.\ Rev.\ Lett.\  {\bf 88}, 041101 (2002).

\bibitem{Keranen}
P.~Keranen, J.~Maalampi and J.~T.~Peltoniemi,
Phys.\ Lett.\ B {\bf 461}, 230 (1999).


\bibitem{peon} 
B.~Peon, ``Is Hinchliffe's Rule True?,''
Print-88-0582 (full text available via SPIRES).

\end{thebibliography}
\end{document}